\documentclass[twocolumn,showpacs,preprintnumbers,superscriptaddress]{revtex4-1}            
\usepackage[T1]{fontenc}   
\usepackage{amsfonts}    
\usepackage{amsmath,amsbsy,amssymb,graphicx}                         
\usepackage{times}
\usepackage{color}

\begin{document}                 
\title{Electromagnon on the Surface of Magnetic Topological Insulator}  
\author{Yusuke~Hama}  
\affiliation{National Institute of Informatics, 2-1-2 Hitotubashi, Chiyoda-ku, Tokyo 101-8430, Japan} 
\author{Naoto~Nagaosa}
\affiliation{RIKEN Center for Emergent Matter Science (CEMS), Wako, Saitama 351-0198, Japan} 
\affiliation{Department of Applied Physics, The University of Tokyo, Bunkyo-ku, Tokyo 113-8656, Japan}
\date{\today}
\begin{abstract}{
We investigate theoretically the electromagnon 
on the surfaces of the magnetic topological insulator thin films.  
It is found that when the magnetic asymmetry between the top and bottom surfaces is there,  the ferromagnetic resonance  is driven by the
electric field which is two orders of magnitude more efficient compared with that by the magnetic field.
The resonant frequency of the electromagnon is also estimated.}  
  
\end{abstract}

\maketitle

\section{Introduction}\label{intro}
Nowadays topological insulator (TI) is one of the central topics in condensed 
matter physics \cite{HansanKane,Qietal1,AndoTIreview,TIbook}.
The two major features are topologically nontrivial bulk band structures 
and  the existence of the gapless surface state whose low-energy effective 
Hamiltonian is described by the spin-momentum locked
Weyl Hamiltonian \cite{Qietal1,Shanetal1,Luetal1}. 
The electromagnetic response owing to the topological bulk band is called topological 
magnetoelectric effect \cite{HansanKane,Qietal1,Qietal2,AndoTIreview,TIbook,Essinetal}. 
It  is described by the axion Lagrangian density
$\mathcal{L}=\theta(\alpha/4\pi^2)\boldsymbol{E}\cdot\boldsymbol{B}$ with
 $\alpha=e^2/\hbar c$ the fine structure constant, $\theta=\pi (0)$ for topologically nontrivial (trivial) insulator.  
The associated electromagnetism is described such that a
magnetization is induced by an electric field as 
$4\pi\boldsymbol{M}=\alpha\boldsymbol{E}$ 
while a magnetic field generates an electric polarization as 
$4\pi\boldsymbol{P}=\alpha\boldsymbol{B}$.
This indicates that the bulk TI exhibits the electromagnetism as multiferroics,
which are the materials where the magnetism and 
ferroelectricity coexist~\cite{reviewmultiferroics1,reviewmultiferroics2,reviewmultiferroics3}. 
 It clearly represents that the 
topological magnetoelectric effect is nothing but  
the linear magnetoelectric effect which is an inherent property of multiferroics.
The multiferroics also have interesting magneto-electric optical properties such
as the emergence of the electromagnon having a natural frequency in the GHz to THz 
regime and directional dichroism \cite{reviewmultiferroics3}. 
Likewise the bulk TI, the surface state of the TI also exibits the physical multi-functionality, e.g.,
spin-charge transport phenomena, the magnetization dynamics in ferromagnet coupled TI, 
and optical phenomena intrinsic to the spin-momentum locking~\cite{Yokoyamaetal2,
Nomuraetal,Garateetal,RaghuetalTIspinplasmon,Tseetal1,HourTIphoto,McIveretalTIphotocurrent,
Tserkovnyaketal1,MellniketalTISTT,ShiomiTISP,Sakaietal,Taguchietal,OgawaetalTIphotocurrent}.
We then naturally expect that these rich TI surface properties enable to realize the 
multiferroic magneto-electric optical phenomena. 

In this paper, we study the magnon of the ferromagnet on the surface of magnetic TI.
Especially, we focus on the thin films of magnetic TI, which are now experimentally 
realized by magnetic doping, for instance,  by Cr, V, and Mn ions 
\cite{Checkelskyetal1,Changetal,Fanetal,Leeetal,Kouetal,Mogietal,MnTIFMR,Vdoping1,Vdoping2,Mogietal2,ChangandLi}, 
or coating the ferromagnetic insulator (FMI) such as EuS to TI (TI/FMI heterosturcture) \cite{ChangandLi,Mooderagroup1}. 
The examples of three dimensional TI materials are $\text{Bi}_2\text{Se}_3$, 
$\text{Bi}_2\text{Te}_3$, their compounds  
$\text{Bi}_2(\text{Se}_x\text{Te}_{1-x})_3$, and $\text{Sb}_2\text{Te}_3$. 
  
We demonstrate that when an electromagnetic field (emf) is applied to the magnetic TI and the 
surface state is in the quantized anomalous Hall state, 
the magnetization couples not only to the magnetic field but also to the electric field so that 
the magnetization behaves as the electric polarization.
In such circumstance, the ferromagnetic resonance (FMR) due to the electric field is much stronger than that by the magnetic field.
This indicates that the magnon as the fluctuation of the electric polarization, i.e., the electromagnon, is induced. 
Such electromagnon can be created when the exchange coupling 
between the magnetization and surface state is large enough and a magnetic 
anisotropy along the out-of surface plane is strong.
We then analyze in what conditions do this electromagnon emerges 
in terms of the helicities of the surface states, magnetization direction, 
the exchange coupling signs, and the helicity of the emf. 
Here we use the terminology ``helicity'' for the TI surface states to describe how the spin and momentum are locked.
Mathematically, the helicity of the TI surface state is defined by the inner product between momentum vector and the vector product of an unit
vector perpendicular to the surface and a spin vector whose components are described by the Pauli matrices. Two TI surface states, top and bottom surface states,
 have opposite helicities.
For emf field we use it to distinguish whether the light is left or right circularly polarized.

We show that the relevant quantities to create the electromagnon are the helicities of the surface states, the magnetization direction, and the emf helicity. 
In other words, the electromagnon emerges when two surfaces have asymmetric magnetization configuration
and the emf applied with the appropriate combination of the emf helicity and magnetization direction.
Such situation can be realized by making the system into a semi-magnetic thin film or by surface doping using two different types of magnetic ions. 
Furthermore, we analyze the resonant frequency of the electromagnon.

\section{Model}\label{model} 
We first focus on one of the surface, say the top surface in the 
two-dimensional plane $\boldsymbol{x}=(x,y)$ located at $z=0$. 
The surface state is coupled to an applied emf
and the magnetization through the exchange coupling. 
The Hamiltonian of this system is
\begin{align}
\mathcal{H} = {  v_{\text{F}}} 
(\sigma^y \Pi_x  -\sigma^x\Pi_y)
- JS{n}_a(\boldsymbol{x},t)\sigma^a,
 \label{eq:hamiltonian1}
\end{align}
where $ v_{\text{F}}$ is the Fermi velocity of the surface state, 
$\sigma^a$ $(a=x,y,z)$ is the Pauli matrices describing the spin,
$\Pi_j=-i\hbar\partial_j+eA_j$ ($j=x,y$) with
$p_j=-i\hbar\partial_j$ the momentum operator, $-e<0$ is the electron charge,
and $A_j$  the external electromagnetic vector potential, respectively.
The summation for the spin index $a$ is taken here.
$J=J^\ast n_0$ is the exchange coupling  with $J^\ast$ being
intrinsic to the ferromagnetic materials and $n_0=d/(a^2a_z)$ is
the averaged two-dimensional sheet density  
of magnetic ions with  $a$ the separation between the localized spin 
(the lattice constant) in the $xy$ plane, $a_z$ the lattice constant for the $z$ axis,
and  $d$ is the interaction range. Here we set $a_z=d.$
The vector $n_a$ is a unit vector for the localized spin field with $S$ its spin magnitude. 
We assume that the ferromagnet has an easy axis parallel to the 
$z$ direction and treat $S_z$ as a constant. 
We analyze the fluctuations of the in-plane magnetization $n_{x,y}$ 
generated by the applied emf.

From Eq. \eqref{eq:hamiltonian1}, one can define the generalized vector
potential as  
\begin{align}
\mathcal{A}_i(\boldsymbol{x},t) = A_i(\boldsymbol{x},t)+a_i(\boldsymbol{x},t),  \qquad(i=x,y)
\label{eq:vector1}
\end{align}
where we have defined the emergent vector potential
\begin{align}
a_x(\boldsymbol{x},t) &=- \frac{JS}{ev_{\text{F}}}n_y(\boldsymbol{x},t), \quad
a_y(\boldsymbol{x},t)  =  \frac{JS}{ev_{\text{F}}}n_x(\boldsymbol{x},t).
\label{eq:vector2}
\end{align}
When the Fermi energy is inside the gap $2\Delta_ {\text{M}}= 2| JS n_z|$ induced by the
$z$-component of the magnetization,  the system shows the quantized 
anomalous Hall effect, and the electromagnetic response of the system is 
characterized by the half-quantized Hall conductance 
$\sigma_H =  \frac{e^2}{2 h}\text{sgn}(J n_z)$ such that the electric current density takes the form
\begin{align}
J_i(\boldsymbol{x},t) &=  \epsilon_{ij}\sigma_H(E_j+e_j),
\label{QHcurrent1}
\end{align}
where $E_j=-\partial_t A^j$ ($=\partial_t A_j$) and $e_j=-\partial_t a^j$ ($=\partial_t a_j$)
are the external and emergent electric fields, respectively. 
$\epsilon_{ij}$ is the antisymmetric tensor with $\epsilon_{xy}=-\epsilon_{yx}=1.$
The above quantized Hall current originates from the Chern-Simons term. 

Now let us focus on the second term of the Hall current in Eq. \eqref{QHcurrent1} which is described as the response to the emergent electric field.
From the relation between the electric polarization and the polarization current $\boldsymbol{P}=\partial\boldsymbol{j}_P/\partial t,$ 
we see that the in-plane magnetization $n_{x,y}$ can be identified with the two-dimensional electric polarization \cite{Nomuraetal} 
\begin{align}
{P}^M_i(\boldsymbol{x},t)=\frac{JS\sigma_H}{ev_{\text{F}}}n_i(\boldsymbol{x},t).
\label{2dpolarization}
\end{align}
Thus, Eq. \eqref {2dpolarization} reflects that the surface of the magnetic TI exhibits both the ferromagnetism and 
the ferroelectricity, i.e., the multiferroicity.
Putting the Fermi velocity  $v_{\text{F}}$=$ 4 \times 10^5$m/sec and $J^\ast$=145meV$\cdot$nm$^2$ \cite{Leeetal},
this electric polarization is estimated as $P^M_i \cong 7.02 \times 10^{-30} n_0 Sn_i$C/m$^3$ when 
the density of the magnetic ions $n_0$ is measured in the unit of m$^{-2}$. 

\section{Ferromagnetic Resonance and Electromagnon}\label{FMREM}
We now study the FMR on the surface of the magnetic TI.
For doing so, we apply both the left and right circular polarized light (LCPL and RCPL).
We show that the resonance due to the electric field is stronger than that by the magnetic 
field implying the electromagnon creation.
We also study the condition for the emergence of this electromagnon. 
The resonant frequency of this magnon is also being estimated.

The magnetization in Eq. \eqref {2dpolarization} couples to the applied emf as
\begin{align}
{H}_{em}&=-n_0\left(P^M_i E_i
+\hbar\gamma_e Sn_iB_i
\right)\notag\\
&=-n_0S\left(\frac{J^\ast\sigma_H}{ev_{\text{F}}}E_i
+\hbar\gamma_eB_i
\right)n_i,
\label{emfcouple}
\end{align}
where $\gamma_e=1.761\times10^{11}$rad/$\text{T}\cdot\text{s}$ 
is the gyromagnetic ratio of the electron. 
Let us estimate the coupling  strength between the electric field and the magnetization for the case of magnetic doping by
using the same values for the exchange coupling $J^\ast$ and the Fermi velocity $v_{\text{F}}$ presented previously.
Then the coupling strength between the electric field and a single localized spin is estimated as 2.11$\times10^{-21} n_0S$J/T$\cdot$m$^2$
where we used the relation $E_{0x}=\pm cB_{0y}$ and $E_{0y}=\mp cB_{0y}$
(the sign - (+) corresponds to the LCPL (RCPL)).
In contrast, the Zeeman interaction strength in Eq. \eqref{emfcouple} is equal to 1.86$\times10^{-23} n_0S$J/T$\cdot$m$^2$.
Thus, we see that the coupling strength due to the electric field is much stronger than that owing to the magnetic field. 
This indicates that when the emf is applied to the surface of the magnetic TI, instead of the ordinary magnon created by the magnetic field
the electromagnon can be dominantly generated due to the nature of the spin-momentum locking. 
Here we note that in Ref. \cite{Henketal}, the magnetic surface gap induced by the out-of plane magnetization has been estimated for Mn-doped TI by the first principle calculation which is 16meV.
The surface gap reported in Ref.  \cite{Leeetal} is around 60meV, and therefore, they are comparable implying that the coupling between the electric field and the magnetization $n_i$  exceeding the Zeeman coupling can also be realized for Mn-doped TI.

Next we study the dynamics of the magnetization. The Zeeman coupling term is going to be neglected
since it is much smaller than the electric-field channel coupling.
When the  transverse emf 
$\boldsymbol{E}(\boldsymbol{x},z,t)=(E_{0x},E_{0y},0)e^{i( kz\mp\omega t)},
\boldsymbol{B}(\boldsymbol{x},z,t)=(B_{0x},B_{0y},0)e^{i( kz\mp\omega t)}$ is applied
($k$ is the wavenumber and $\omega$ the dispersion satisfying  $\omega=c k$ with $c=3.0\times10^8$m/s 
the speed of light in the vacuum. The minus (positive) sign in the plane wave
describes the LCPL (RCPL)), 
the equation of motion for the magentization $n_i$ is given by
\begin{align}
\dot{n}_+-i\omega_0 n_+=i\frac{JeE_+}{4\pi\hbar v_{\text{F}}SC_1}\text{sgn}(J n_z), \label{eom1}
\end{align}
where the dot ``$\cdot$'' represents the time derivative, $n_+\equiv n_x+in_y$, $E_+\equiv E_x+iE_y$, and 
\begin{align}
\omega_0&=\frac{Ka^2a_z n_0}{C_1}, \label{naturalfrequency}\\
C_1&=\frac{\hbar n_0n_z}{S}+\frac{J^2}{4\pi\hbar v^2_{\text{F}}}\text{sgn}(J n_z), \label{coefficient}
\end{align}
with $K$ a magnetic anisotropy constant.
Equation of motion \eqref{eom1} describes a forced oscillation of the in-plane magnetization $n_i$ 
where the sum between the Berry phase term and anisotropic energy term play a role of oscillator 
with a resonant frequency given by Eq. \eqref{naturalfrequency}, while the Chern-Simons term act as a time dependent external force in terms of the electric field. 
By using the physical parameters shown in the previous discussion, we have
$\hbar n_0 n_z/S\simeq 1.29\times10^{-16}\cdot(n_z/S)$J$\cdot$s/m$^2$ and $J^2/4\pi\hbar  v^2_{\text{F}}\simeq 3.80\times10^{-18}$J$\cdot$s/m$^2$.
Here we have used proximate values of data for lattice constants of tetradymites as  $a$=$4.2\times10^{1/3}\simeq9.05$\AA, and 
$a_z$=30$\times10^{1/3}\simeq6.46\times10$\AA   
 \cite{TIcrystaldata} with assuming 10\% Cr-concentration. 
Thus, the coefficient $C_1$ in Eq. \eqref{coefficient} is approximated as 
$C_1\approx \hbar n_0n_z/S$, and subsequently, for the frequency \eqref{naturalfrequency} we obtain $\omega_0\approx Ka^2a_z S/\hbar n_z$.
The LCPL is expressed by $E_+=E_0e^{-i\omega t}$ and we take the electomagnetic field amplitude  $E_{0}$ as a time-independent quantity. 
By assuming the solution for Eq. \eqref{eom1} in a form $n_+=\bar{n}e^{- i\omega t}$ ($\bar{n}$ is time independent) we have the solution
\begin{align}
n^{\text{t}}_{+,\text{L}}=\frac{C^{\text{t}}_2 e^{-i\omega t}}{ v_{\text{F}}(\omega+\omega^{\text{t}}_0)}, \quad \text{with} \quad
C^{\text{t}}_2=-\frac{|J|e E_0}{4\pi\hbar^2 n_0 |n_z|},
\label{eomsolutions1}
\end{align}
where we introduced the superscript t and the subscript L representing the top surface and the left circular polarization, respectively.
The solution for the RCPL can be obtained by the replacement $\omega\rightarrow-\omega.$
Saying the external frequency $\omega>0$, as we see from  the approximated form of the resonant frequency shown above,
whether the resonance on the surface magnetic TI occurs 
or not depends on the emf helicity of the light and the magnetization direction: the FMR by the electric field is triggered by the LCPL (RCPL)
 when the magnetization is pointing toward negative (positive) $z$ direction.

Next, let us estimate the resonant frequency of the electromagnon $f_0=\omega_0/2\pi$. 
In Ref. \cite{Fanetal}, the out-of plane anisotropic magnetic field $B_K$ and the saturation magnetization $M_S$ were measured in the Cr-doped TI thin films as $B_K$=0.9T and $M_S$=16$\times10^3$J/T$\cdot$m$^3$. By using the relation $K$=$B_KM_S/2$,
the magnetic anisotropic constant becomes  $K$=7.2$\times10^3$J/m$^3$,
and  the resonant frequency is estimated as $f_0\approx 5.75\times10\cdot (S/n_z)$GHz.

Usually the magnetic impurities are doped in the bulk TI uniformly, and the magnetization is also realized for the  bottom surface \cite{Checkelskyetal1,Changetal,Fanetal,Leeetal,Kouetal,Mogietal,MnTIFMR,Vdoping1,Vdoping2,Mogietal2,ChangandLi}.  
Thus, we need to discuss whether the FMR is triggered or not by the sum between the top and bottom magnetization, 
and indeed, this is what we observe in the experiment.
The above argument for the top surface can be similarly applied to the analysis for the bottom surface. 
The magnetization for the bottom surface corresponding to Eq. \eqref{eomsolutions1} is obtained by first $v_{\text{F}}\rightarrow-v_{\text{F}}$ 
since the helicities of top and bottom surfaces are oppposite. Second, the resonant frequency is in the order of 10GHz, and thus, we have
$kd_z=\omega d_z/c< 10^{-5}\ll1,$ and the plane-wave part of the emf on the bottom surface can be approximated as 
$e^{i(kd_z\mp\omega t)}\approx e^{\mp i\omega t}.$ Here we took the thickness of the TI thin film  $d_z=100$\AA.
Then the magnetization on the bottom induced by the emf is obtained by replacing the quantitiessuch as
 magnetic anisotropy energy, the exchange coupling, for the bottom surface ones.
The total in-plane magnetization $n_{i}=n^\text{t}_{i}+n^\text{f}_{i}$ for the LCPL becomes
\begin{align}
n_{+,\text{L}}=\left[\frac{C^{\text{t}}_2}{\omega+\omega^{\text{t}}_0}-\frac{C^{\text{b}}_2}{\omega+\omega^{\text{b}}_0}\right]
\frac{e^{-i\omega t}}{v_{\text{F}}},\label{eomsolutions2}
\end{align}
 while  $n_{+,\text{R}}$ is obtained by the replacement $\omega\rightarrow-\omega$. 

Let us study the conditions for the electric-field induced surface FMR based on Eq. \eqref{eomsolutions2}.
First, consider the case when the single type of magnetic ions are dopped to the whole TI sample uniformly. 
Then the same magnetization is generated for the top and bottom surfaces where we have  $C^{\text{t}}_2=C^{\text{b}}_2$ and $\omega^{\text{t}}_0=\omega^{\text{b}}_0$.
In this case, the cancellation of the magnetization occurs owing to the opposite helicities between the top and bottom surface states,   
and thus, the FMR and the associated electromagnon are not generated.
To avoid this circumstances and generate the FMR and the electromagnon,
 we next consider the different ways of magnetic doping. 
One way is to dope the magnetic ions only to the single surface, i.e., the semi-magnetic thin film configuration \cite{Mogietal}. 
Then one of the surfaces shows the FMR described by Eq. \eqref{eomsolutions1}. 
Another way is to perform the magnetic doping using two types of ions, for instance, doping V ions to the top surface while Cr ions to the bottom surface \cite{Mogietal2}. 
In such circumstance, since $C^{\text{t}}_2\neq C^{\text{b}}_2$ and $\omega^{\text{t}}_0\neq \omega^{\text{b}}_0,$ there is no magnetization cancellation between the top and 
bottom surfaces, and we will observe the FMR as well as the electromagnon. 

We also examine the case for  TI/FMI heterosturcture settings.
When EuS, EuO, or YIG are used as FMI, it seems to be difficult to realize the FMR and the associated electromagnon
because the magnetic anisotropy is along the in-plane direction \cite{ChangandLi,Mooderagroup1} and the exchange coupling might be weak.
Instead of them, using MnSe might be a good choice. It has been investigated by the first-principle calculation that the surface gap about 54meV can be induced with 
Mn ion having an easy axis along the out-of plane \cite{Luoetal}, while in Ref. \cite{Eremeevetal} the gap of the Weyl surface state is 8.5meV.
Both values are comparable with the magnetic surface gap obtained in \cite{Leeetal} which is around 60meV. 
Although the electronic states and the associated ferromagnetism at the interface between the FMI and TI are quite complicated in TI/FMI heterosturctured systems,
we still believe that the electromagnon on the surface TI can also be created  for these systems  with such choice.

Consequently, in order to generate the electric-field induced FMR and electromagnon, we first have to prepare the asymmetric magnetization configurations between
two surfaces so as to prevent from the cancellation originating from the opposite  TI surface-state helicities. This can be done by either the magnetic doping or coating the FMI on one of the
surfaces of TI. Once such magnetization configurations are setup, the electromagnon emerges with the proper combinations of the magnetization direction and emf helicity.

\section{Conclusion}\label{CON}
 In this paper we have investigated the FMR on the surface of TI.
We have found that when the exchange coupling is large enough around 10meV, the magnetization behaves as two-dimensional electric polarization and 
couples to the electric field so that such coupling exceeds the Zeeman coupling. This reflects that the surface of the magnetic TI exhibits the multiferroics.
As a result, the electric-field induced FMR and the associated electromagnon as a fluctuation of the two-dimensional electric polarization is generated. 

By deriving the equation of motion \eqref{eom1} and its solution \eqref{eomsolutions2},  we have carefully analyzed in what conditions does the electromagnon emerges
by focusing on the  magnetization configuration, the exchange-coupling sign, the surface-state helicities, and the  emf helicity. 
We demonstrated that to generate the electromagnon we have to create the asymmetry between the top and bottom surfaces.
 This could be done, for instance, by doping the magnetic ions only to the single surface or using two different kinds of ions.
Besides the magnetic doping, this electric-field induced FMR and the associated electromagnon can also be realized by the TI/FMI heterostructure
with the proper choice of FMI.
Finally, we have estimated the resonant frequency of the electromagnon  for a Cr-dopping case with the 10\% concentration, which was in the order of 10 GHz. 
By increasing the magnetic anisotropic energy or the saturation magnetization about one or two orders than the above case,
we may create the electromagnon in the THz regime, implying that we can perform the ultrafast manipulation of the 
magnetization by the electric field owing to the spin-momentum locking of the surface TI. 

We note that in Ref. \cite{Garateetal}, the Hall-current (or the electric field)  induced magnetization switching, i.e. the inverse spin-Galvanic effect,
and the associated magnon was pointed out. Further, in Ref. \cite{Sakaietal} the effect on conductivity due to the electric-field induced magnon has been presented.
Although the electric-field induced magnon has been referred to in these articles, in our paper we have found some new insights.
 First we have numerically shown that the coupling between the electric field and magnetization mediated by the TI surface state is stronger than that between the magnetic field and magnetization, i.e. the possibilty of the electric-field induced FMR. For doing this we have used the recent experimental data and compared with the results due to the first principle caluculation. 
 Second, we then have focused on two surfaces and examined in what conditions in terms of the magnetization configuration and its direction
 as well as the surface-state  and emf-field helicities does the electromagnon emerge on TI surfaces. 
Our result indicates not only the posibility of the electromagnon emergence but may guide to explore the hidden electromagnetic and optical properties of surfaces of TI as multiferroics, or those of other two-dimensional multiferroic materials.  
 
Recently, the FMR has been performed in magnetic TI \cite{MnTIFMR,ShiomiTISP}.
The condition adopted in these studies, however, are different from ours:
The magnetic TI in Ref. \cite{ShiomiTISP} consists of permalloy (Ni$_{81}$Fe$_{19}$) and TIs where the surface states are in the metallic regime.
In Ref. \cite{MnTIFMR}, the magnetization induced by the Mn-doping has an easy axis paralell to the surface-plane alignment.
Once the conditions for the magnetization and TI surface states proposed in our paper are applied experimentally,  
the electromagnon may be observed by the FMR in the near future.

\acknowledgements
Y.~H thanks Makiko Nio, Ryutaro Yoshimi, Kenji Yasuda, and Masataka Mogi for fruitful discussion and comments. 
This work was supported by RIKEN Special Postdoctoral Researcher Program (Y.~H) and 
by JSPS Grant-in-Aid for Scientific Research (No. 24224009, and No. 26103006) from MEXT, 
Japan (N.~N).

\end{document}